\documentclass[runningheads]{llncs}
\usepackage[T1]{fontenc}
\usepackage{graphicx}
\usepackage{color}
\usepackage{float}
\usepackage{graphicx}
\usepackage{amsmath}
\usepackage{mathtools}
\usepackage{bm}
\usepackage{hyperref}
\usepackage[capitalise]{cleveref}

\usepackage{float}
\usepackage{multirow}
\usepackage{adjustbox}
\restylefloat{table}
\usepackage{algorithm}
\usepackage{algpseudocode}
\graphicspath{./images/}

\newcommand{\bra}[1]{{\left\langle #1 \right|}}
\newcommand{\ket}[1]{{\left| #1 \right\rangle}}

\newcommand{\ori}{\vee}
\newcommand{\andi}{\wedge}
\newcommand{\noti}{\overline}
\newcommand{\xori}{\oplus}

\usepackage{quantikz}

\definecolor{AW}{HTML}{286D8C}

\usepackage{tikz} 
\usetikzlibrary{matrix,positioning}
\usepackage{amsfonts}
\usepackage{amssymb}
\usepackage{cite}

\algnewcommand{\To}{\textbf{To }}
\algnewcommand\Input{\item[\textbf{Input:}]}%
\algnewcommand\Output{\item[\textbf{Output:}]}%

\definecolor{rpLightPurple}{HTML}{232136}
\definecolor{rpDarkPurple}{HTML}{1F1D2E}
\definecolor{rpWhite}{HTML}{e0def4}
\definecolor{rpPink}{HTML}{c4a7e7}
\definecolor{rpBlue}{HTML}{3e8fb0}
\definecolor{rpLightBlue}{HTML}{9ccfd8}
\definecolor{rpYellow}{HTML}{f6c177}
\definecolor{rpRed}{HTML}{eb6f92}
\definecolor{rpGreen}{HTML}{31748f}
\definecolor{rpBlack}{HTML}{26233a}
\definecolor{rpOtherBlack}{HTML}{6e6a86}
\definecolor{rpMagenta}{HTML}{c4a7e7}
\definecolor{amathyst}{HTML}{9063CD}
\definecolor{pink}{HTML}{E75480}
\usepackage{subfig}

\tikzstyle{block} = [draw=rpPink, rounded corners, inner xsep=3pt, inner ysep=3pt, fill=rpWhite, text centered]
\tikzstyle{arrow} = [thick, ->, >=Stealth]
\usepackage{quantikz}
\usetikzlibrary{shapes.geometric, arrows.meta, positioning}

\hypersetup{
   colorlinks = true,
    urlcolor=blue
}
\usepackage[normalem]{ulem}
\usepackage{lipsum}

\graphicspath{{figures/}}
\usepackage{tikz-network}
\usetikzlibrary{fit, shapes.arrows}
\usetikzlibrary{shapes.geometric, arrows}
\usetikzlibrary{positioning,chains,fit,shapes,calc}

\tikzstyle{boxblue} = [rectangle, rounded corners, minimum width=3cm, minimum height=1cm,text centered, draw=black, fill=blue!30]
\tikzstyle{boxred} = [rectangle, rounded corners, minimum width=3cm, minimum height=1cm,text centered, draw=black, fill=red!30]
\tikzstyle{boxgreen} = [rectangle, rounded corners, minimum width=3cm, minimum height=1cm,text centered, draw=black, fill=green!30]
\tikzstyle{io} = [trapezium, trapezium left angle=70, trapezium right angle=110, minimum width=3cm, minimum height=1cm, text centered, draw=black, fill=orange!30]
\tikzstyle{process} = [rectangle, minimum width=3cm, minimum height=1cm, text centered, draw=black, fill=orange!30]
\tikzstyle{decision} = [diamond, minimum width=3cm, minimum height=1cm, text centered, draw=black, fill=green!30]
\tikzstyle{arrow} = [thick,->,>=stealth]

\tikzstyle{decision} = [diamond, draw, fill=blue!20, 
    text width=4.5em, text badly centered, node distance=3cm, inner sep=0pt]
\tikzstyle{block} = [rectangle, draw, fill=blue!20, 
    text width=5em, text centered, rounded corners, minimum height=4em]
\tikzstyle{line} = [draw, -latex']
\tikzstyle{cloud} = [draw, ellipse,fill=red!20, node distance=3cm,
    minimum height=2em]

\usepackage{float}

\usepackage{array, booktabs, makecell, multirow}

\usepackage{amsmath,calc}

\definecolor{graphPurple}{RGB}{185,75,185}

\DeclareMathOperator*{\argmin}{argmin}

\begin{document}
\title{An Exclusive-Sum-of-Products Pipeline for QAOA}
%
%
\author{Matthew Brunet \inst{1,3}\orcidID{0009-0009-0966-3947} 
\and
Shilpi Shah\inst{2}\orcidID{0009-0004-7962-7515} 
\and
Mostafa Atallah \inst{3, 4}\orcidID{0009-0004-8187-6932} \and
Anthony Wilkie \inst{3}\orcidID{0009-0007-9808-1759}  \and
Rebekah Herrman \inst{3}\orcidID{0000-0001-6944-4206}
}
\authorrunning{M. Brunet et al.}
%
\institute{Winthrop University, Rock Hill SC 29733, USA \and
Rutgers University, New Brunswick NJ 08901, USA
\\
\and
University of Tennessee Knoxville, Knoxville TN 37996, USA\\
\email{rherrma2@utk.edu}
\and
Cairo University, Giza 12613, Egypt}
\maketitle              
\begin{abstract}
The quantum approximate optimization algorithm is commonly used to solve 
combinatorial optimization problems.
While unconstrained problems map naturally into the algorithm, incorporating constraints typically requires penalizing constraint violations in the objective function. 
In this work, we propose an alternative approach that encodes constraints as Boolean expressions in exclusive-sum-of-products (ESOP) form before penalization. 
We test this method on the maximum independent set problem using graphs with 3 to 20 vertices and find that ESOP constraint formulations achieve higher approximation ratios than standard constraint penalization methods, with percent increases of up to 30.3\%. Furthermore, ESOP constraint formulations result in higher approximation ratios than standard QAOA penalization approaches after one layer of the algorithm on approximately 64\% of the tested graphs.

\end{abstract}

\section{Introduction}\label{sec:intro}
The quantum approximate optimization algorithm (QAOA) \cite{Farhi2014FQAOA} is a variational quantum algorithm that has the potential to provide a speed-up over classical algorithms for solving combinatorial optimization problems \cite{montanez2024towards, shaydulin2024evidence, tate2023warm}. 
Recent QAOA research has resulted in numerous variations of the algorithm that seek to either decrease the circuit depth or problem size \cite{ponce2025graph, zhou2023qaoa}, allow for more expressible circuits \cite{herrman2022multi, wang2020xy, wilkie2024quantum, vijendran2024expressive}, or reduce the space of feasible solutions \cite{golden2021threshold, wilkie2025learning, tate2020bridging}.

While unconstrained combinatorial optimization problems such as MaxCut can easily be mapped into QAOA \cite{Ozaeta_2022}, solving constrained combinatorial optimization problems, such as the Maximum Independent Set (MIS) problem, with the QAOA is more challenging. 
One of the main methods for handling constraints is to penalize them in the objective function, turning the problem into an unconstrained one that can be naturally mapped to QAOA \cite{herrman2021globally, lucas2014ising}.
These approaches can require ancillary qubits and can involve squaring sums of constraints, leading to products of several variables.
This penalization can lead to complex cost Hamiltonians whose optimization landscape is difficult to navigate due to barren plateaus and local minima \cite{kazi2024lie, larocca2025barren, allcock2024dynamical}.
Another method of handling constraints is to encode the constraints directly into the initial state of the QAOA circuit, reducing the search space to only the feasible states \cite{bartschi2020grover, wang2020xy, golden2021threshold, wilkie2025learning, tate2020bridging, goldstein2024convergence}.
The drawback to this method is the difficulty in creating an efficient (in circuit depth or number of two-qubit gates) state preparation circuit to encode the feasible states.
For single constraints, such as linear equality constraints \cite{Bartschi_2022, wang2020xy}, the state preparation can be efficient, however it is not clear if this is easy for general sets of constraints, specifically constraints that share large amounts of variables.
Other approaches for constrained optimization for QAOA try to avoid penalization altogether \cite{angara2025art, bucher2025if}.

The foundational work by Hadfield et al. \cite{hadfield2019quantum} first established systematic methods for encoding Boolean constraints into phase oracles and Hamiltonians for quantum algorithms.
Building on this foundation, recent approaches such as Boolean-Hamiltonian-Transformation QAOA (BHT-QAOA) \cite{al2024bht} have extended these techniques to generate QAOA cost Hamiltonians by marking bit-strings that do not satisfy Boolean constraint expressions.
In contrast to other penalization methods, BHT-QAOA formulates each constraint as an exclusive-sum-of-product (ESOP) Boolean expression.
Interestingly, ESOP expressions have been used in quantum circuit compilation algorithms to reduce T gate count \cite{meuli2019evaluating, schmitt2019scaling} and in quantum walk assisted QAOA (QW-QAOA) to solve constrained combinatorial optimization problems \cite{marsh2019quantum}, though the QW-QAOA approach updates the mixing Hamiltonian to search only over feasible solutions.

In this work, we first introduce background on QAOA, MIS, and ESOPs in Sec.~\ref{sec:background}.
Then, we describe how to use ESOP expressions to encode constraints into QAOA, using MIS as a test problem, in Sec.~\ref{sec:method}.
In Sec.~\ref{sec:results}, we compare the ESOP constraint approach to standard QAOA penalty approaches to solve MIS on over 3000 graphs with between 3 and 20 vertices and find that encoding constraints using the ESOP formulation results in up to a $30.3\%$ percent increase in approximation ratio over the standard QAOA constraint encoding approach.
Finally, we conclude with a discussion in Sec.~\ref{sec:discussion}.

\section{Background}\label{sec:background}
In this section, we provide background information for QAOA, the MIS problem, SOPs, and ESOPs.

\subsection{QAOA}
QAOA is a variational quantum algorithm in which two unitaries, a cost unitary $U(C, \gamma)= e^{-i C \gamma}$ and a mixing unitary $U(B, \beta)= e^{-i B \beta }$, are applied to an initial state $\ket{s}$.
The cost Hamiltonian, $C$, is a diagonal matrix that encodes the combinatorial optimization problem $f(x)$ being solved.
The construction of $C$ for the MIS problem will be described in more detail later.
The mixer Hamiltonian $B$ allows for probability amplitude to be transferred between the computational basis states.
In order for QAOA to converge to the optimal solution to the combinatorial optimization problem, the initial state $\ket{s}$ should be an eigenstate of $B$.
$B$ is often chosen to be the transverse field mixer, $B = \sum_i \sigma_i^x$, and $\ket{s} = \ket{+}^{\otimes n}$. 
The final state after $p$ iterations of the algorithm is
\begin{equation*}
    \ket{\gamma, \beta} = U(B, \beta_p)U(C, \gamma_p) \cdots U(B, \beta_1)U(C, \gamma_1)\ket{s}.
\end{equation*}
The real-valued parameters $\beta$ and $\gamma$, often called angles, are selected to minimize the expected value of the cost Hamiltonian $C$,
\begin{equation*}
    \gamma^*, \beta^* = \argmin_{\gamma, \beta} \langle C \rangle = \argmin_{\gamma, \beta} \bra{\gamma, \beta} C \ket{\gamma, \beta}.
\end{equation*}
Classical optimization routines such as BFGS \cite{NumericalRecipesBFGS} and COBYLA \cite{powell1994direct} are commonly used to find the angles.

A common measure of success for QAOA is the approximation ratio (AR), which for constrained optimization, is defined as
\begin{equation*}
    \frac{\langle C \rangle - C_{\mathrm{max}}}{C_{\mathrm{min}} - C_{\mathrm{max}}}
\end{equation*}
where $C_{\mathrm{max}}$ ($C_{\mathrm{min}}$) is the largest (smallest) eigenvalue of $C$.
The AR is a real number between 0 and 1, with values closer to 1 indicating better solutions to the combinatorial optimization problem.


\subsection{MIS}\label{sec:mis}

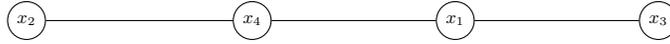
\begin{figure}
\centering
\begin{tikzpicture}[scale=3]
\begin{scope}[every node/.style={scale=.75,circle,draw}]
\node (A) at (-1.9,0) {$x_2$};
\node (B) at (-0.9,0) {$x_4$};
\node (C) at (0,0) {$x_1$}; 
\node (D) at (0.9,0) {$x_3$}; 
\end{scope}

\draw  (A) -- (B);
\draw  (B) -- (C);
\draw  (C) -- (D);

\end{tikzpicture}
\caption{Path on four vertices, $P_4$. This has three unique maximum independent sets: $\{x_1, x_2\}$, $\{x_2, x_3\}$, and $\{x_3, x_4\}$.}
\label{fig:misexample}
\end{figure}

The goal of the MIS problem on a simple graph $G= (V,E)$ is to find the largest subset of vertices $I \subset V$ such that for all $i,j \in I$, $ij \notin E$. For an example, see Fig.~\ref{fig:misexample}. 
We can formulate the MIS problem as an integer program
\begin{align*}
    \max_{x \in \{0, 1\}^{|V|}}& \sum_{i \in V} x_i \\
    \text{s.t.}&\; x_i + x_j \leq 1, \quad \forall (i, j) \in E.
\end{align*}

The MIS problem is commonly written as a quadratic unconstrained binary optimization (QUBO) problem of the form \cite{xu2025qaoa, pichler2018quantum, ebadi2022quantum, brady2024iterative}
\begin{equation*}
  \min \;\;\;  -\sum_{i \in V} x_i + J\sum_{ij \in E} x_ix_j,
\end{equation*}
which is mapped to $H_C$ using the substitution $x_i \rightarrow \frac{I - Z_i}{2}$ \cite{hadfield2021representation}.
Note that this formulation recasts MIS as a minimization problem by negating the objective function. 
Throughout this work, we shall refer to this encoding as the ``standard QAOA constraint encoding" method.

\subsection{SOPs and ESOPs}\label{sec:esop}
Boolean expressions are commonly written in a SOP formulation \cite{abhyankar2009minimal, arevalo1978method, luccio2002new}, in which products of Boolean variables (ANDs) are added together (ORs).
Any Boolean expression can be converted to an equivalent ESOP, which is a Boolean expression consisting solely of exclusive sums of products of literals that can be negated.
One can convert a Boolean expression to an ESOP expression using the identity
\begin{align*}
    a \vee b &= a \noti{b} \oplus b = a \xori b \xori ab.
\end{align*}

ESOP formulations of Boolean expressions have applications in machine learning \cite{perkowski1995application} and Toffoli gate synthesis \cite{sanaee2010esop}.
While converting Boolean expressions to ESOP expressions is hard, there exist parallel algorithms for converting an arbitrary Boolean expression to an ESOP formulation \cite{papakonstantinou2014parallel}.
Furthermore, there exist efficient algorithms for minimizing ESOP expressions \cite{stergiou2004fast}.
The author of \cite{hadfield2021representation} introduced rules for converting Boolean expressions into sums and products of Pauli-$Z$ operators, which are summarized in Table~\ref{tab:bol_hams}.

\begin{table}[]
    \centering
    \[
    \renewcommand{\arraystretch}{1.5}
    \begin{array}{|c|c||c|c|}
    \hline 
    f(x) & H_f & f(x) & H_f \\
    \hline 
    \hline 
    x & \frac{1}{2} I - \frac{1}{2} Z & \noti{x} & \frac{1}{2} I + \frac{1}{2} Z \\
    \hline 
    x_1 \oplus x_2 & \frac{1}{2} I - \frac{1}{2} Z_1 Z_2 & \bigoplus_{j=1}^k x_j & \frac{1}{2} I - \frac{1}{2} Z_1 Z_2 \cdots Z_k \\
    \hline
    x_1 \andi x_2 & \frac{1}{4} I - \frac{1}{4} \left( Z_1 + Z_2 - Z_1 Z_2\right) & \bigwedge_{j=1}^k x_j & \frac{1}{2^k} \prod_j \left(I - Z_j\right) \\
    \hline
    x_1 \ori x_2 & \frac{3}{4} I - \frac{1}{4} \left( Z_1 + Z_2 + Z_1 Z_2\right) & \bigvee_{j=1}^k x_j & I - \frac{1}{2^k} \prod_j \left(I + Z_j\right) \\
    \hline
    \noti{x_1} \noti{x_2} & \frac{3}{4} I + \frac{1}{4} \left( Z_1 + Z_2 - Z_1 Z_2\right) & x_1 \implies x_2 & \frac{3}{4} I + \frac{1}{4} \left( Z_1 - Z_2 + Z_1 Z_2\right) \\
    \hline
    \end{array}
    \]
    \caption{The table from \cite{hadfield2021representation} that shows how to write a Boolean expression $f(x)$ as a Hamiltonian $H_f$.}
    \label{tab:bol_hams}
\end{table}

Another identity within \cite{hadfield2021representation} that will be commonly used in this work is that for two Boolean expressions $f$ and $g$ represented by Hamiltonians $H_f$ and $H_g$ respectively, then 
\begin{equation}\label{eq:Hamiltonian_addition}
    H_{f \oplus g} = H_f + H_g - 2H_fH_g.
\end{equation}
Note in particular that for a Boolean expression of the form $(x_i \wedge x_j) \oplus (\noti{x_j} \wedge x_k)$, where $f = (x_i \wedge x_j)$ and $g = (\noti{x_j} \wedge x_k)$, then $H_f = \frac{1}{4}(I - Z_i)(I - Z_j)$ and $H_g = \frac{1}{4}(I + Z_j)(I - Z_k)$.
Substituting these into Eq.~\ref{eq:Hamiltonian_addition} yields
{\scriptsize
\begin{align*}
  H_{f \oplus g}& =  \frac{1}{4}(I - Z_i)(I - Z_j) + \frac{1}{4}(I + Z_j)(I - Z_k) - 2\frac{1}{4}(I - Z_i)(I - Z_j)\frac{1}{4}(I + Z_j)(I - Z_k) \notag \\
  & = \frac{1}{4}(I - Z_i)(I - Z_j) + \frac{1}{4}(I + Z_j)(I - Z_k) - \frac{1}{8}(I - Z_i)(I - Z_j)(I + Z_j)(I - Z_k) \notag \\
  & = \frac{1}{4}(I - Z_i)(I - Z_j) + \frac{1}{4}(I + Z_j)(I - Z_k) \notag \\
  &= H_f + H_g\label{eq:simplification} 
\end{align*}
}%
as $(I - Z_j)(I + Z_j) = 0$.
Thus, one can eliminate higher-order terms from the Hamiltonian formulation of $f \oplus g$ if there exists a variable of the form $x_j$ in one expression and $\noti{x_j}$ in the other. 
In general, suppose we have a set of Boolean expressions $\{f_i\}_i$ where $f_i = \bigwedge_{u \in V_i} \ell_{\sigma(u)}(x_u)^i$, $V_i \subset V$, $\sigma(u) \in \{0,1\}$, and
\[
\ell_{\sigma(u)}^i = \begin{cases}
    x_u & \sigma(u) = 0, \\
    \noti{x_u} & \sigma(u) = 1
\end{cases}.
\]
If $|\{u \in V: \exists i, j \text{ s.t. } \ell_{\sigma(u)}^i(x_u) \neq \ell_{\sigma(u)}^j(x_u)\}| > 1$, that is, if there is at least one vertex for which $x_u$ is in $f_i$ and $\noti{x_u}$ is in $f_j$, then
\begin{align}
    H_{\bigoplus_{i} f_i} = \sum_i H_{f_i}. \label{eq:general_simp}
\end{align}

\section{ESOP pipeline}\label{sec:method}


The following sections describe the framework for generating the QAOA cost Hamiltonian, $H_C$, for the MIS problem.
In this framework, we first convert each constraint in the constrained optimization problem to a Boolean expression.
Then, we rewrite the Boolean expression into an ESOP formulation.
The ESOP formulation is then multiplied by a large penalty parameter and added to the objective function of the optimization problem.
We shall refer to this as the "ESOP pipeline".
The objective function is then converted to a cost Hamiltonian, $H_C$, via the rules from \cite{hadfield2021representation}.
Once $H_C$ has been generated, we then use it in QAOA with the standard Pauli-$X$ mixer.
To automate and test the procedure, we developed a Python library (found at \cite{ESOPGitHub}) which takes a Networkx graph as input and produces a QAOA circuit for the MIS problem on said graph.
These circuits are then simulated using the Qiskit Aer simulator \cite{QiskitAer} in Qiskit version 1.3.2 \cite{Qiskit}.
Classical optimization of angles $\gamma$ and $\beta$ is employed using COBYLA \cite{powell1994direct}.

Note that the methods in this section apply to any constrained optimization problem, however we use MIS as an illustrative example.
The rest of this section will apply the ESOP pipeline to MIS problems.

\subsection{MIS Boolean encoding}

The first step of the ESOP pipeline for a given graph $G$ is to generate a Boolean function $v_{\mathrm{IS}}(\overrightarrow{x_G}):\{0,1\}^n \rightarrow \{0,1\}$ satisfying
\begin{align*}
v_{\mathrm{IS}}(\overrightarrow{x_G})&= \begin{cases}
    1 & \text{if } \{k : \overrightarrow{x_G}[k]=1\} \text{ form an independet set in G} \\
    0 & \text{otherwise}.
\end{cases}
\end{align*}
To do this, we will require a Boolean function that compares incident vertices.
The independence constraints for the MIS problem can be written as
\begin{equation*}
\bigwedge\limits_{ij \in E}^{|E|}\noti{x_{i}\andi x_{j}},
\end{equation*}
since if both $x_i$ and $x_j$ are 1, the expression is equal to $0$, signaling that both $i$ and $j$ cannot be included in $\overrightarrow{x_G}$.
Using De'Morgans law, this is equivalent to
\begin{equation}\label{eq:DNF}
     v_{\mathrm{MIS}}(\overrightarrow{x_G}) = \noti{\bigvee\limits_{ij\in E}^{|E|}{x_{i}\andi x_{j}}}.
\end{equation}
This formulation of the constraints can be refactored into an ESOP using algebraic techniques and then added to the objective function.

\subsection{Converting Boolean constraints to ESOP}
By inverting both sides of Eq.~\eqref{eq:DNF}, notice that

\begin{equation*}
\noti{v_{\mathrm{IS}}(\overrightarrow{x_G})} = \bigvee\limits_{ij\in E} {x_{i}\andi x_{j}}.
\end{equation*}

For simplicity, we rewrite each $x_i \andi x_j$ as some $e_k$ for $0 \leq k \leq |E|-1 $.
Then for all edges $e_k \in E$,
\begin{align*}
\noti{v_{\mathrm{IS}}(\overrightarrow{x_G})} &= \bigvee\limits_{k=0}^{|E|-1} {e_{k}} \\
                   &= e_{0} \ori e_{2} \ori \cdots \ori e_{|E|-1}.
\end{align*}

Applying the identity $a \vee b = (a \wedge \neg{b}) \oplus b$ three times yields
\begin{align*}
\noti{v_{\mathrm{MIS}}(\overrightarrow{x_G})} &=( (e_{0} \andi \noti{e_{1}}) \xori  e_{1}) \ori e_{2} \ori e_3 \ori \cdots \ori e_{|E|-1} \\
                    &= (e_{0} \andi \noti{e_{1}} \andi \noti{e_{2}}) \xori  (e_{1}\andi \noti{e_{2}}) \xori (e_2) \ori e_3 \\
                    &\quad \ori \cdots \ori \ e_{|E|-1} \\
                    &= (e_{0} \andi \noti{e_{1}} \andi \noti{e_{2}} \andi \noti{e_3})  \xori  (e_{1}\andi \noti{e_{2}}\andi \noti{e_3}) \xori (e_2 \andi \noti{e_3}) \\
                    &\quad \ori \cdots \ori \ e_{|E|-1}.
\end{align*}
It is clear that continuing this process until there are no more OR operations will yield an equivalent expression of $v_{\mathrm{IS}}(\overrightarrow{x_G})$, written only using the XOR, AND, NOT, operations:

\begin{align*}
     \noti{v_{\mathrm{IS}}(\overrightarrow{x_G})} &= e_0\andi\noti{{e_1}}\andi\noti{e_2}\andi \noti{e_3} \andi \noti{e_4}\andi \cdots \andi \noti{e_{|E|-1}} \\
    & \quad \oplus \ {e_1}\andi\noti{e_2}\andi \noti{e_3} \andi \noti{e_4} \andi \cdots \andi \noti{e_{|E|-1}} \\
    & \quad \oplus \ {e_2}\andi\noti{e_3}\andi \noti{e_4}\andi \cdots \andi \noti{e_{|E|-1}} \\
    & \quad \oplus \ {e_3}\andi \noti{e_4} 
    \andi \cdots \andi \noti{e_{|E|-1}} \\
    & \quad \xori \cdots \xori \ e_{|E| - 2} \andi \noti{e_{|E|-1}} \oplus e_{|E|-1} 
\end{align*}
This expression can be written more succinctly as
\begin{align*}
     \noti{v_{\mathrm{IS}}(\overrightarrow{x_G})} &= \left[ \bigoplus_{k=0}^{ |E| - 2}e_k \bigwedge_{r=k+1}^{|E|-1} \noti{e_{r}} \right]  \xori e_{|E|-1} \notag \\
    &= \left[ \bigoplus_{k=0}^{|E|-2} (x_{i_k} \andi x_{j_k}) \bigwedge_{r=k+1}^{|E|-1} \noti{(x_{i_r} \andi x_{j_r})} \right] \xori (x_{i_{|E|-1}} \andi x_{j_{|E|-1}}) \\
    &= \left[ \bigoplus_{k=0}^{|E|-2} (x_{i_k} \andi x_{j_k}) \bigwedge_{r=k+1}^{|E|-1} (\noti{x_{i_r}} \ori \noti{x_{j_r}}) \right] \xori (x_{i_{|E|-1}} \andi x_{j_{|E|-1}}) \\
    &= \left[ \bigoplus_{k=0}^{|E|-2} (x_{i_k} \andi x_{j_k}) \bigwedge_{r=k+1}^{|E|-1} \left[(\noti{x_{i_r}} \andi x_{j_r}) \xori \noti{x_{j_r}}\right] \right] \xori (x_{i_{|E|-1}} \andi x_{j_{|E|-1}}).
\end{align*}

The above process transforms the individual products into even longer terms of conjunctions and negated conjunctions that can be simplified with De'Morgan's law.
See Sec.~\ref{sec:example} for an example of the ESOP derived for the path on four vertices.

\subsection{Transformation from Boolean expression to a Hamiltonian}

Once the constraints have been converted into the ESOP formulation $\bigoplus_i c_i$, where $c_i = \bigwedge_j x_j$ with some $x_j$ being negated, we will convert each term into a Hamiltonian using Table~\ref{tab:bol_hams}.
Specifically, each variable $x_j \rightarrow \frac{1}{2} (I - Z_j)$ and each $\noti{x_j} \to \frac{1}{2} (I + Z_j)$.
Where we diverge from Table~\ref{tab:bol_hams} is how we handle $\bigwedge_{j=1}^k x_j$.
Similar to \cite{al2024bht}, we will multiply every second variable in the term $c_i$ by $-1$, resulting in
\begin{align*}
    c_i = \bigwedge_{j = 1}^k x_j \to H_{c_i} =  \frac{1}{2^k} \prod_j (-1)^{j+1}(I - Z_j).
\end{align*}
Since there exist $x_{j_r}$ and $\noti{x_{j_r}}$ terms in the ESOP formulation, we can use Eq.~\ref{eq:general_simp} to write the constraint penalty Hamiltonian as
\begin{align*}
    H_c:= H_{\oplus_i c_i} = \sum_i H_{c_i}.
\end{align*}
With the MIS objective Hamiltonian $H_{\mathrm{MAX}} = \sum_{j \in V} \frac{1}{2} (I - Z_j)$, the full cost Hamiltonian is given by
\begin{align*}
    H_C = - H_{\mathrm{MAX}} - 2|V| H_c = -H_{\mathrm{MAX}} - 2|V| \sum_{i} H_{c_i}.
\end{align*}
Note that normally when minimizing an objective function, one adds the Hamiltonians representing constraints to the objective.
However, since we derived $\noti{v_{\mathrm{IS}}(\overrightarrow{x_G})}$ above, we subtract it from the objective instead of adding it.
Also note that we use $2|V|$ as a penalty in our simulations, but any sufficiently large penalty parameter suffices.

\subsection{Example: $P_4$}\label{sec:example}
Consider the path graph on four vertices, denoted $P_4$, in Fig.~\ref{fig:misexample}.
We will apply the ESOP pipeline to this graph to find a cost Hamiltonian that encodes MIS on this problem.
Then, we simulate QAOA solving this problem using the Qiskit Aer simulator.
It is clear that there are three solutions to this problem: $\{x_1, x_2\}$, $\{x_2, x_3\}$, and $\{x_3, x_4\}$.
Furthermore, the objective of this problem is $\max \sum_{i=1}^4 x_i$, or equivalently, $\min \; -\sum_{i=1}^4 x_i$.

The ESOP that encodes the MIS constraints is
   \begin{align*}
   \noti{v_{\mathrm{IS}}(\overrightarrow{x_{P_4}})} &= (x_2 \andi x_4) \ori (x_4 \andi x_1) \ori (x_1 \andi x_3) \\
   &= e_1 \ori e_2 \ori e_3 \\  
   &= (e_1 \andi \noti{e_2} \andi \noti{e_3}) \xori (e_2 \andi \noti{e_3}) \xori (e_3) \\
   &= (x_2 \andi x_4 \andi \noti{x_4 \andi x_1} \andi \noti{x_1 \andi x_3}) \xori (x_4 \andi x_1 \andi \noti{x_1 \andi x_3}) \xori (x_1 \andi x_3)\\
   &= (\noti{x_1} \andi x_2 \andi {x_4}) \xori (x_1 \andi \noti{x_3} \andi {x_4}) \xori (x_1 \andi x_3).
   \end{align*}

We shall refer to $(\noti{x_1} \andi x_2 \andi {x_4})$ as $c_1$, $(x_1 \andi \noti{x_3} \andi {x_4}) $ as $c_2$, and $(x_1 \andi x_3)$ as $c_3$.
Note that by Eq.~\ref{eq:general_simp}, $H_{c_1 \oplus c_2 \oplus c_3} = H_{c_1} + H_{c_2} + H_{c_3}$. Thus,
\begin{equation*}
    H_C = -H_{\mathrm{MAX}} - 8 \sum_{i=1}^3{H_{c_i}}
\end{equation*}
where
   \begin{align*}
       &H_{c_1} = \frac{1}{8}(I + Z_{1})(-I + Z_2)(I-Z_4) \\
       &H_{c_2} = \frac{1}{8}(I-Z_1)(-I-Z_3)(I-Z_4)\\
       &H_{c_3} = -\frac{1}{4}I + \frac{1}{4}(Z_1 + Z_3 - Z_1Z_3).
   \end{align*}

\begin{figure}
    \centering
    \includegraphics[width = 0.75\textwidth]{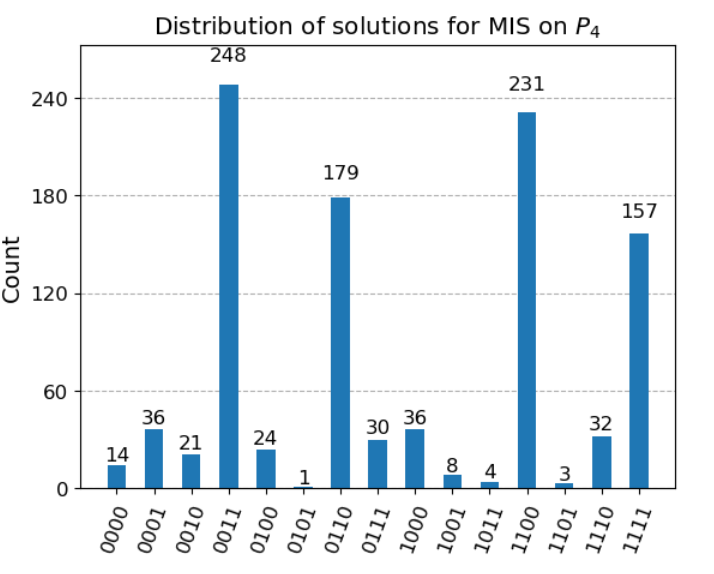}
    \caption{Measurements for solving MIS on $P_4$ with only one layer of QAOA and constraints encoded as ESOPs using 1024 shots. The rightmost qubit corresponds to the value of $x_1$ and the leftmost corresponds to $x_4$.
    } 
    \label{fig:p1_for_4_node}
\end{figure}

Fig.~\ref{fig:p1_for_4_node} shows the counts when simulating QAOA with this cost Hamiltonian.
Note that the optimal solutions of $\{x_1, x_2\}$, $\{x_2, x_3\}$, and $\{x_3, x_4\}$ correspond to bitstrings $1100$, $0110$, and $0011$ respectively.

\section{Results}\label{sec:results}

We implement the ESOP pipeline for penalizing constraints with QAOA on over 3000 graphs: all 3-vertex (2 instances), 4-vertex (6 instances), 5-vertex (21 instances), and 6-vertex graphs graphs (112 instances), 700 randomly sampled 7-vertex graphs, 500 randomly generated graphs for all vertex counts with 8 vertices and between 11 and 14 vertices, and 50 randomly generated graphs for all vertex counts between 15 and 20 vertices.
All tested graphs are connected and nonisomorphic.
Note that graphs with less than 8 vertices were generated using the \texttt{g6} files from \cite{mckay}.
We calculate the approximation ratio for each instance and then compare the results to the standard QAOA penalization technique described in Sec.~\ref{sec:mis}.

\begin{figure}
    \centering
    \begin{subfloat}
        \centering
        \includegraphics[height=6cm]{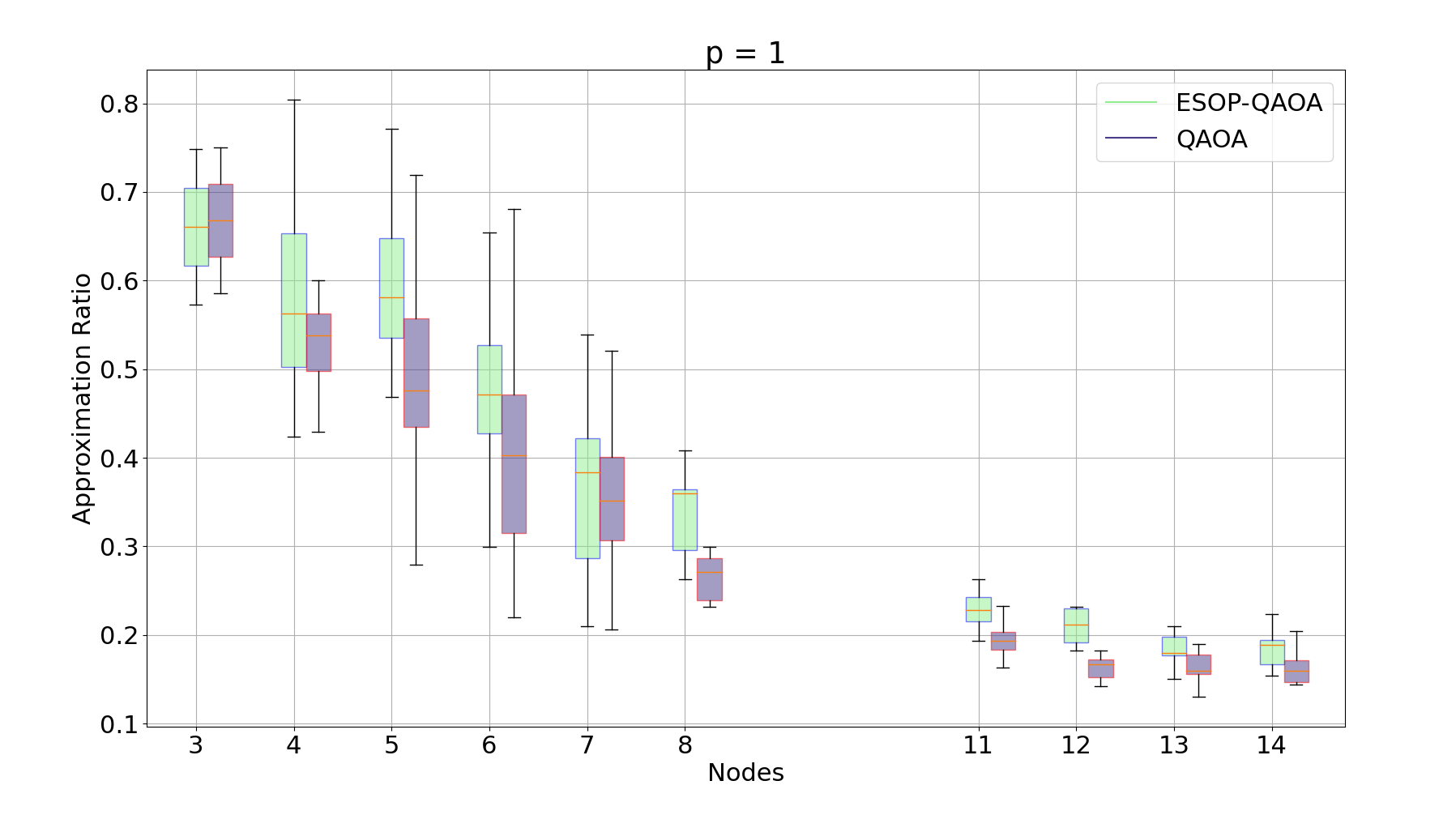}
    \end{subfloat}
    
    \begin{subfloat}
        \centering
        \includegraphics[height=6cm]{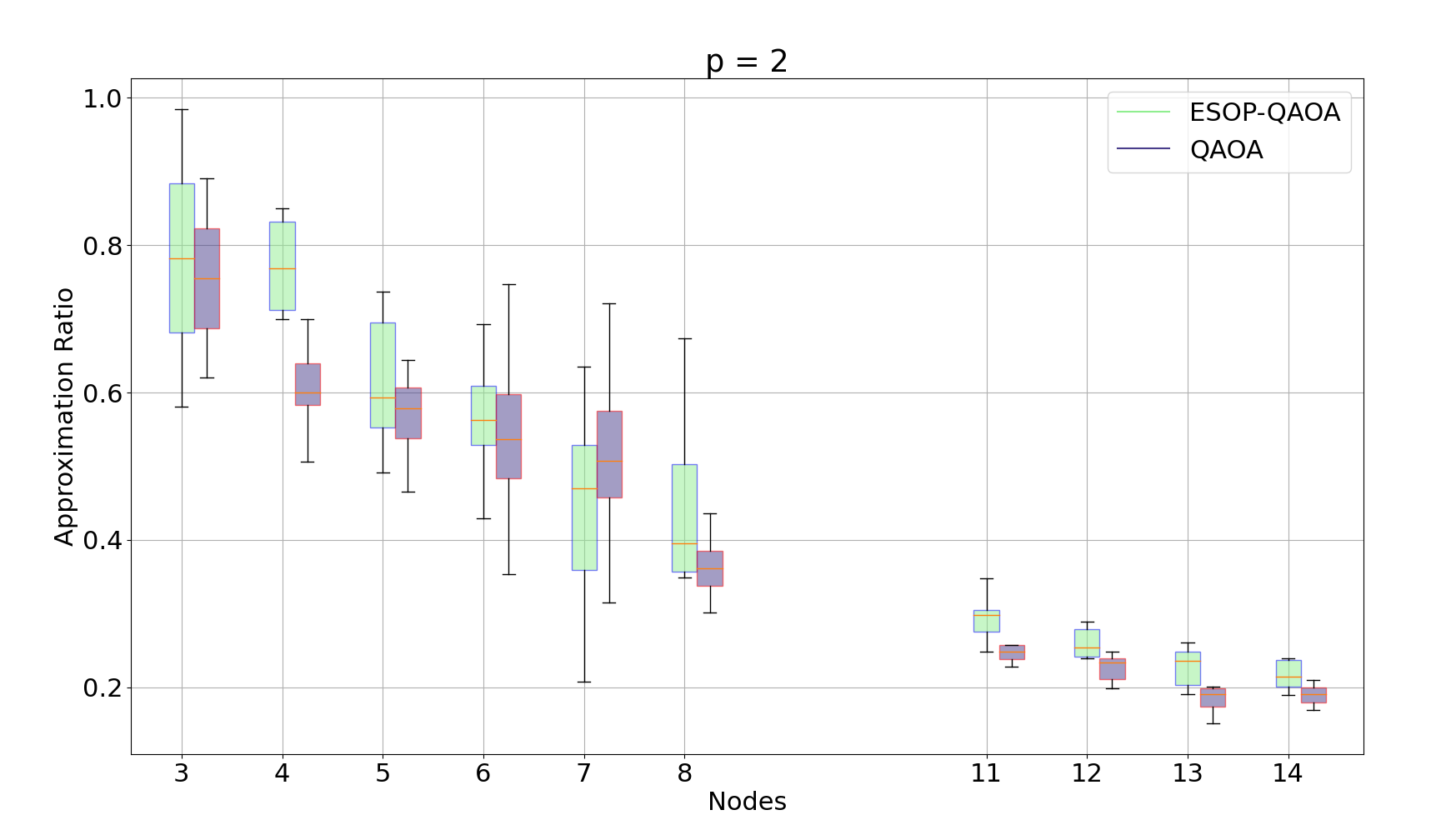}
    \end{subfloat}
    
    \begin{subfloat}
        \centering
        \includegraphics[height=6cm]{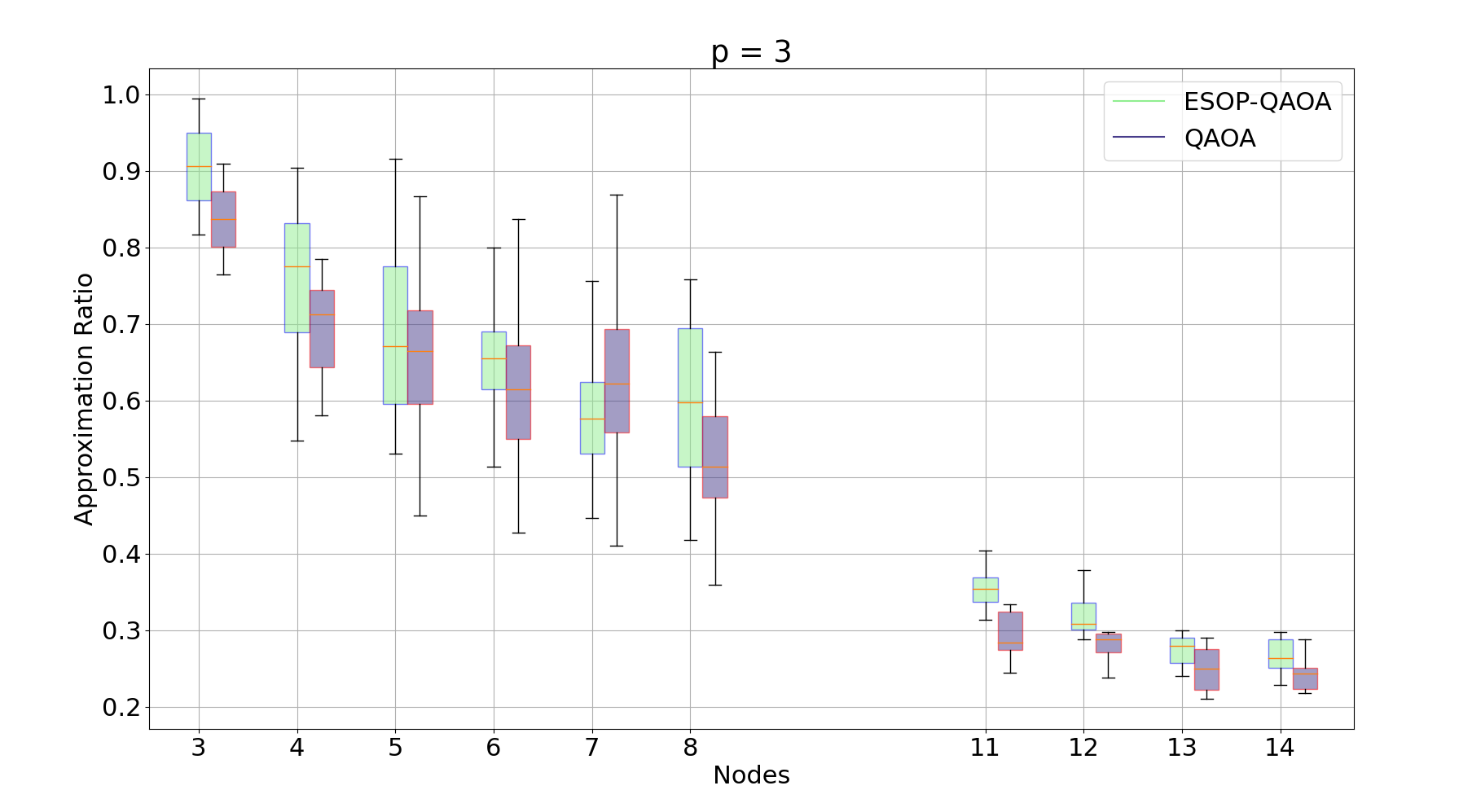}
    \end{subfloat}
    \caption{Comparison of approximation ratios between regular QAOA and our ESOP-QAOA for the MIS problem graphs of size 3 up to 14. Note we did not test our method of graphs of size 9 or 10 vertices.}
    \label{fig:p1_3_to_14}
\end{figure}

\begin{figure}
    \centering
    \begin{subfloat}
        \centering
        \includegraphics[height=6cm]{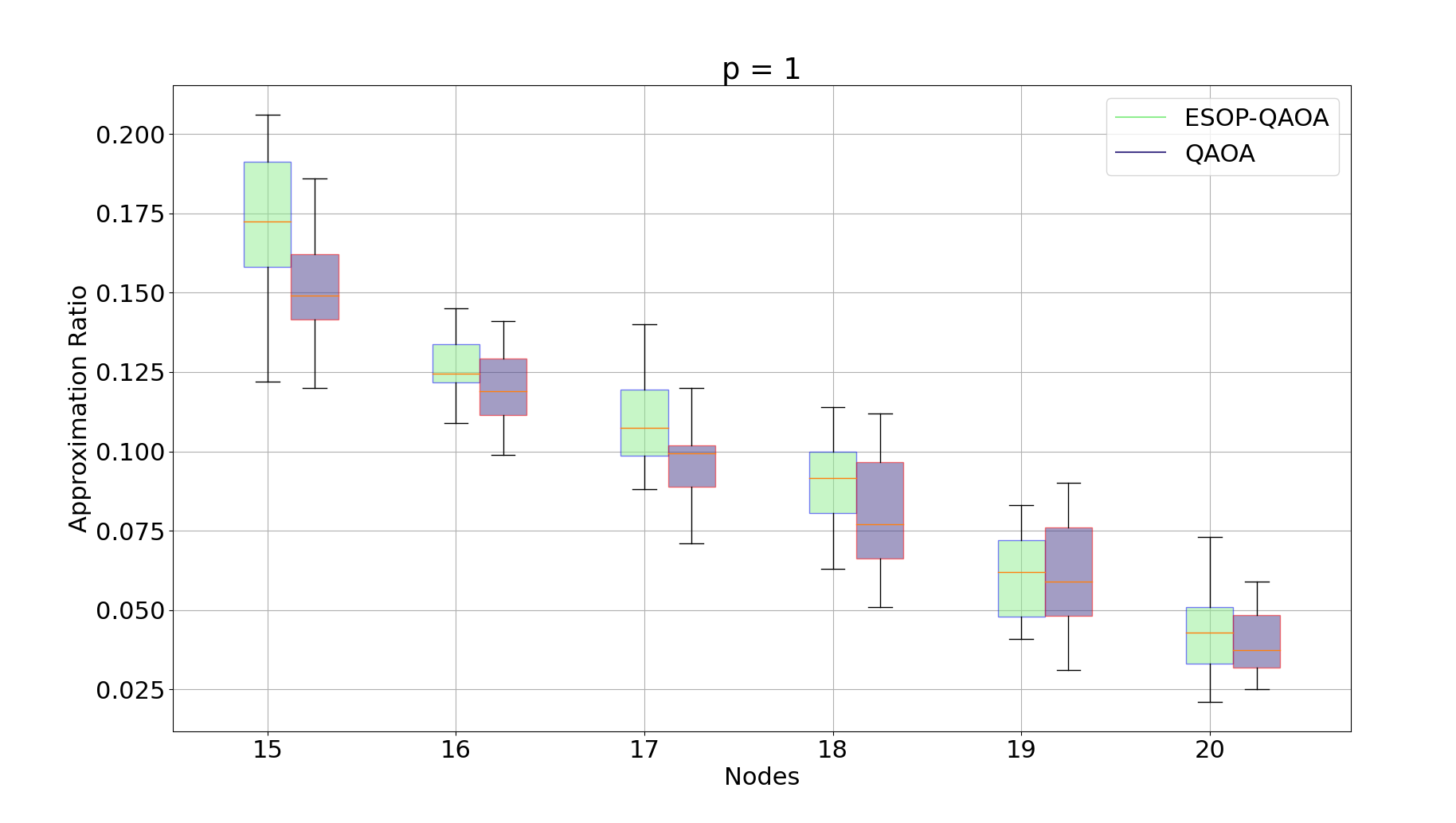}
    \end{subfloat}
    
    \begin{subfloat}
        \centering
        \includegraphics[height=6cm]{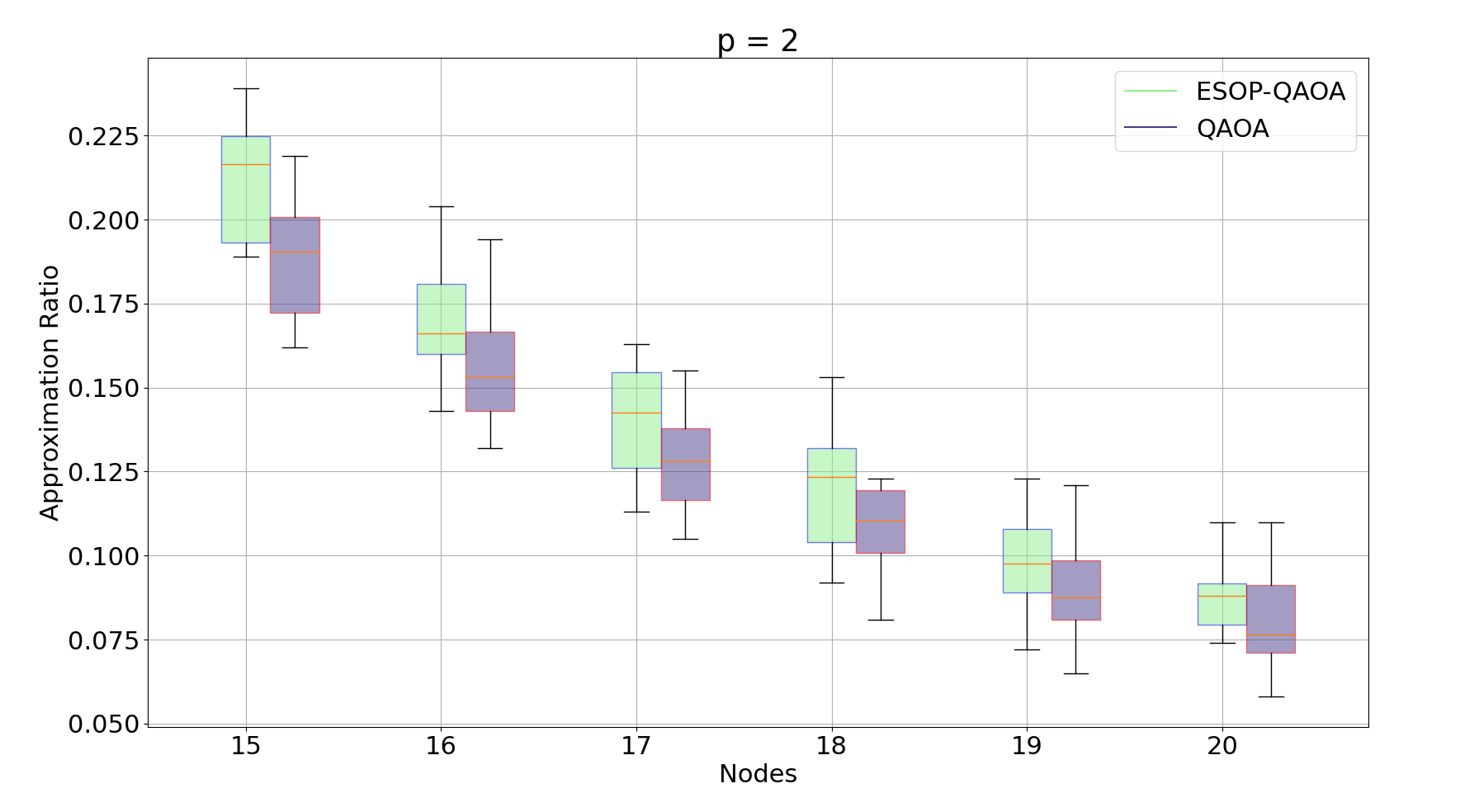}
    \end{subfloat}
    
    \begin{subfloat}
        \centering
        \includegraphics[height=6cm]{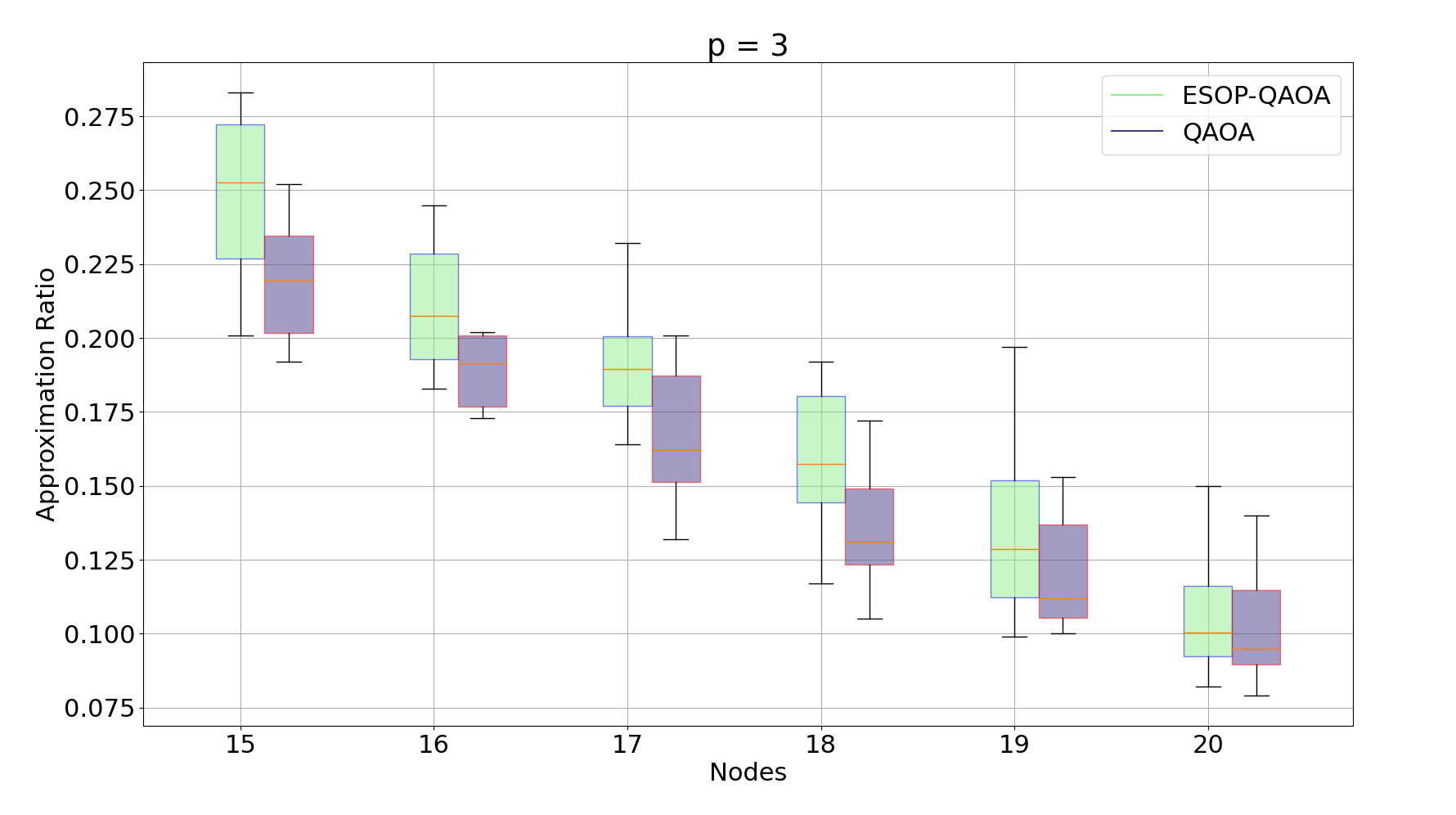}
    \end{subfloat}
    \caption{Comparison of approximation ratios between regular QAOA and our ESOP-QAOA for the MIS problem graphs of size 15 up to 20.}
    \label{fig:p1_15_to_20}
\end{figure}

The approximation ratios for tested graphs with between 3 and 14 vertices is found in Fig.~\ref{fig:p1_3_to_14}, while the approximation ratios for graphs with between 15 and 20 vertices is found in Fig.~\ref{fig:p1_15_to_20}.
Note that when $p=1$, the ESOP constraint encoding approach results in higher average approximation ratios across each graph size except $n=3$, which is a dataset consisting of only two graphs.
At $p=2$ and $p=3$, the ESOP constraint encoding method results in higher approximation ratios than QAOA for all size graphs except $n=7$.
The average approximation ratios of both approaches are listed in Table~\ref{tab:approximation_ratios_improved}, as well as the percent change between the standard MIS constraint encoding for QAOA and the ESOP constraint encoding.  When $p=1$, the ESOP constraint formulations result in higher approximation ratios than standard QAOA penalization approaches on approximately 64\% of the tested graphs.

\begin{table}[]
    \centering
    \small
    \adjustbox{scale=.875}{
    \begin{tabular}{|c|ccc|ccc|ccc|}
    \hline 
     \multirow{2}{*}{\makecell{Vertices}} & \multicolumn{3}{c|}{QAOA p=1} & \multicolumn{3}{c|}{QAOA p=2} & \multicolumn{3}{c|}{QAOA p=3} \\
     \cline{2-10}
     & Standard & ESOP & (\% change) & Standard & ESOP & (\% change) & Standard & ESOP & (\% change) \\
    \hline 
    3 & 0.660 & 0.655 & (\textcolor{red}{-0.8}) & 0.749 & 0.792 & (\textcolor{blue}{+5.7}) & 0.835 & 0.910 & (\textcolor{blue}{+9.0}) \\
    \hline 
    4 & 0.535 & 0.560 & (\textcolor{blue}{+4.7}) & 0.600 & 0.782 & (\textcolor{blue}{+30.3}) & 0.719 & 0.787 & (\textcolor{blue}{+9.5}) \\
    \hline
    5 & 0.470 & 0.580 & (\textcolor{blue}{+23.4}) & 0.592 & 0.597 & (\textcolor{blue}{+0.8}) & 0.655 & 0.658 & (\textcolor{blue}{+0.5}) \\
    \hline
    6 & 0.411 & 0.468 & (\textcolor{blue}{+13.9}) & 0.563 & 0.570 & (\textcolor{blue}{+1.2}) & 0.621 & 0.654 & (\textcolor{blue}{+5.3}) \\
    \hline
    7 & 0.350 & 0.384 & (\textcolor{blue}{+9.7}) & 0.450 & 0.441 & (\textcolor{red}{-2.0}) & 0.625 & 0.586 & (\textcolor{red}{-6.2}) \\
    \hline
    8 & 0.279 & 0.359 & (\textcolor{blue}{+28.7}) & 0.392 & 0.398 & (\textcolor{blue}{+1.5}) & 0.526 & 0.599 & (\textcolor{blue}{+13.9}) \\
    \hline
    11 & 0.194 & 0.233 & (\textcolor{blue}{+20.1}) & 0.231 & 0.250 & (\textcolor{blue}{+8.2}) & 0.292 & 0.350 & (\textcolor{blue}{+19.9}) \\
    \hline
    12 & 0.169 & 0.215 & (\textcolor{blue}{+27.2}) & 0.222 & 0.230 & (\textcolor{blue}{+3.6}) & 0.295 & 0.303 & (\textcolor{blue}{+2.7}) \\
    \hline
    13 & 0.161 & 0.182 & (\textcolor{blue}{+13.0}) & 0.195 & 0.219 & (\textcolor{blue}{+12.3}) & 0.254 & 0.280 & (\textcolor{blue}{+10.2}) \\
    \hline
    14 & 0.156 & 0.191 & (\textcolor{blue}{+22.4}) & 0.194 & 0.211 & (\textcolor{blue}{+8.8}) & 0.245 & 0.255 & (\textcolor{blue}{+4.1}) \\
    \hline
    15 & 0.149 & 0.172 & (\textcolor{blue}{+15.4}) & 0.189 & 0.216 & (\textcolor{blue}{+14.3}) & 0.215 & 0.252 & (\textcolor{blue}{+17.2}) \\
    \hline
    16 & 0.118 & 0.124 & (\textcolor{blue}{+5.1}) & 0.154 & 0.166 & (\textcolor{blue}{+7.8}) & 0.191 & 0.208 & (\textcolor{blue}{+8.9}) \\
    \hline
    17 & 0.098 & 0.112 & (\textcolor{blue}{+14.3}) & 0.130 & 0.141 & (\textcolor{blue}{+8.5}) & 0.160 & 0.190 & (\textcolor{blue}{+18.8}) \\
    \hline
    18 & 0.077 & 0.089 & (\textcolor{blue}{+15.6}) & 0.110 & 0.123 & (\textcolor{blue}{+11.8}) & 0.132 & 0.159 & (\textcolor{blue}{+20.5}) \\
    \hline
    19 & 0.060 & 0.064 & (\textcolor{blue}{+6.7}) & 0.089 & 0.096 & (\textcolor{blue}{+7.9}) & 0.110 & 0.129 & (\textcolor{blue}{+17.3}) \\
    \hline
    20 & 0.036 & 0.041 & (\textcolor{blue}{+13.9}) & 0.077 & 0.092 & (\textcolor{blue}{+19.5}) & 0.095 & 0.100 & (\textcolor{blue}{+5.3}) \\
    \hline
    \end{tabular}}
    \caption{Average approximation ratios comparing standard QAOA penalization with ESOP-QAOA across different circuit QAOA layers $(p=1,2,3)$ and graph sizes. Percentage changes show ESOP performance relative to standard QAOA, with improvements in \textcolor{blue}{blue} and negative percent changes in \textcolor{red}{red}.}
    \label{tab:approximation_ratios_improved}
\end{table}

\section{Discussion}\label{sec:discussion}
In this work, we develop an ESOP pipeline for encoding constrained optimization problems into QAOA.
In this pipeline, we first write all constraints as Boolean expressions, and then write the Boolean expressions into an ESOP formulation, in contrast to other QAOA constraint encoding methods.
Then, we multiply these terms by a large penalty parameter and add it to the objective function.
Our experimental results demonstrate that rewriting constraints as ESOP statements and penalizing them in the QAOA cost Hamiltonian results in significantly higher approximation ratios than standard constraint penalization methods.
Out of 54 test configurations spanning different graph sizes (3-20 vertices) and QAOA layer depths ($p=1,2,3$), ESOP-QAOA achieves higher average approximation ratios in 51 cases.
The method shows particularly large approximation ratio improvements on smaller graphs, with notable increases of $30.3\%$ for $4$-vertex graphs at $p=2$, $28.7\%$ for $8$-vertex graphs at $p=1$, and $27.2\%$ for $12$-vertex graphs at $p=1$.
Even for larger problem instances (15-20 vertices), the ESOP constraint encoding method shows positive increases in approximation ratios, suggesting scalability.

This work highlights the importance of problem encoding in the quantum approximate optimization algorithm, as the ESOP formulation of constraints implements the same logic as the usual Boolean implementation.
The ESOP formulation appears to work well for MIS because of the existence of $x_i$ and $\noti{x_i}$ terms in the Boolean expression, which allows for constraint Hamiltonian simplifications.
Future interesting research avenues include developing problem-based ESOP simplification methods, similar to the higher order terms canceling in MIS, for other constrained optimization problems.
Other future avenues of research include identifying problem classes where ESOP constraints guarantee simpler implementations with shorter circuit depth and fewer ancilla qubits than standard penalty methods, and developing parameter transfer techniques \cite{montanez2025transfer} to further enhance optimization efficiency. Finally, the ESOP constraint formulations result in higher approximation ratios than standard constraint encoding methods on approximately 64\% of the tested graphs. Characterizing these graphs would allow one to select the best encoding method for a particular MIS instance.

\subsubsection*{Code and Data Availability}
\label{sec:codeAndDataAvail}
The code and data for this research can be found at \url{https://github.com/mxttbrunet/Quantum-Walk-Project}.

\begin{credits}
\subsubsection{\ackname} 
M.B., S.S., and R.H. acknowledge NSF CNS-2244512. A.W. and R.H. acknowledge NSF CCF 2210063.
\subsubsection{\discintname}
The authors have no competing interests to declare that are relevant to the content of this article. 
\end{credits}
%
%
%
 \bibliographystyle{splncs04}
 \bibliography{ESOP}
\end{document}